\documentclass[twocolumn,showpacs,floatfix,superscriptaddress,amsmath,amssymb,prl]{revtex4}
\usepackage{mathrsfs}
\usepackage{txfonts}
\usepackage{amssymb}
\usepackage{graphicx}
\usepackage{hyperref}
\usepackage{ulem}
 \usepackage{overpic}
 \usepackage{psfrag}
\usepackage{tabularx}
\usepackage{array}
\usepackage{placeins}

\makeatletter
\renewcommand\subsection{\@startsection
{subsection}{2}{0mm}
 {-\baselineskip}
 {0.5\baselineskip}
{\FloatBarrier\normalfont\Large\bfseries}}
\makeatother
\newcommand{\be}{\begin{equation}}
\newcommand{\ee}{\end{equation}}

\newcommand{\PreserveBackslash}[1]{\let\temp=\\#1\let\\=\temp}

\begin{document}
\title{Universal construction of order parameters for translation-invariant quantum lattice systems with symmetry-breaking order}

\author{Jin-Hua Liu}
\affiliation{Centre for Modern Physics and Department of Physics,
Chongqing University, Chongqing 400044, The People's Republic of
China}
\author{Qian-Qian Shi}
\affiliation{Centre for Modern Physics and Department of Physics,
Chongqing University, Chongqing 400044, The People's Republic of
China}
\author{Hong-Lei Wang}
\affiliation{Centre for Modern Physics and Department of Physics,
Chongqing University, Chongqing 400044, The People's Republic of
China}
\author{Jon Links}
\affiliation{Centre for Mathematical Physics,
School of Mathematics and Physics,
The University of Queensland, 4072, Australia}
\author{Huan-Qiang Zhou}
\affiliation{Centre for Modern Physics and Department of Physics,
Chongqing University, Chongqing 400044, The People's Republic of
China}

\begin{abstract}
For any translation-invariant quantum lattice system with a symmetry group
$G$,  we propose a practical and universal construction of order parameters which identify quantum phase transitions with symmetry-breaking
order. They are defined in terms of the
fidelity between a ground state and
its symmetry-transformed counterpart, and are computed through tensor network representations of the ground-state wavefunction.
To illustrate our scheme, we consider three quantum
systems on an infinite lattice in one spatial dimension, namely,
the quantum Ising model in a transverse magnetic field, the quantum spin-1/2
XYX model in an external magnetic field, and the quantum spin-1 XXZ model
with single-ion anisotropy. All these models have symmetry group $\mathbb{Z}_2$ and exhibit broken-symmetry phases.
We also discuss the role of the order parameters in identifying factorized states.

\end{abstract}


\maketitle

{\it Introduction.} In the conventional Landau-Ginzburg-Wilson
paradigm a basic notion is spontaneous symmetry breaking,
which is traditionally characterized in terms of a local order
parameter~\cite{b4, b5}. Typically, such a local order parameter is
model-dependent and not always obvious to define. Although attempts have been made
to derive a local order parameter
for quantum lattice systems undergoing quantum phase
transitions (QPTs)~\cite{b1,b2}, it is highly desirable to find a
simple, unifying, and model-independent way to characterize symmetry-breaking order
when it occurs in a quantum lattice many-body system.

In this Letter, we address this issue from a quantum information
perspective through the notion of {\it fidelity}. In
Refs.~\cite{b6,zhou,zhou1,b7}, it has been argued that the
ground-state fidelity per site may be used to detect QPTs. Since
this argument is based solely on a basic postulate of quantum
mechanics regarding quantum measurement, the approach is applicable
to quantum lattice systems in any number of spatial dimensions,
regardless of the type of internal order present in quantum
many-body states. It has been confirmed that the ground-state
fidelity per site is able to identify QPTs arising from spontaneous
symmetry breaking~\cite{b6,zhou,zhou1,b7,b3}, the
Kosterlitz-Thouless transition~\cite{b9}, and topological QPTs in
the Kitaev model~\cite{b8}. In an extension of this notion, here we
propose a universal approach to define and compute order parameters
for any translation-invariant quantum lattice system, with a
symmetry group $G$, undergoing a QPT with symmetry-breaking order.
We perform the explicit computation of order parameters through the
use of tensor network algorithms, in particular the matrix product
states (MPS)~\cite{b10,b11, b12} for systems in one spatial
dimension.

To illustrate our scheme, we investigate the following models on an
infinite lattice in one spatial dimension: the quantum Ising model
in a transverse magnetic field, the quantum spin-1/2 XYX model in an
external magnetic field, and the quantum spin-1 XXZ model with
single-ion anisotropy.  All these systems possess a discrete
symmetry group $\mathbb{Z}_{2}$, which is spontaneously broken as
the system undergoes a QPT. Although these examples are restricted
to  $\mathbb{Z}_{2}$-systems on an infinite-size lattice in one
spatial dimension, we emphasize that the scheme extends to any
translation-invariant quantum lattice system, with arbitrary
symmetry group, in any spatial dimension. For systems in higher
spatial dimensions
 the computation of the ground state is accommodated by the tensor product states
(TPS)~\cite{tps}, or equivalently, the projected entangled-pair
states (PEPS)~\cite{peps}.

The construction of these order parameters not only allows us
to locate critical points, but also enables us to identify
factorized states $|\psi(\lambda_f) \rangle$, where $\lambda_f$ is
the so-called factorizing field~\cite{mc,factorized}. Such a
factorized state $|\Psi(\lambda_f) \rangle$ can occur in the
symmetry-broken phase. The fact that no entanglement exists makes it
the most ordered state, with a salient feature that the order parameters
take their maximum value at $\lambda=\lambda_f$.

{\it Universal construction of order parameters for
translation-invariant quantum lattice systems with symmetry-breaking
order.} Consider an infinite-lattice translation-invariant quantum
system with symmetry group $\tilde{G}$ and Hamiltonian $H(\lambda)$, with
$\lambda$ a control parameter. According to Wigner's theorem the representations of
$g\in \tilde{G}$ are either unitary or anti-unitary. The group elements which are represented unitarily form a subgroup $G\subseteq\tilde{G}$, and we hereafter restrict our considerations to this subgroup. Suppose a symmetry-breaking QPT
occurs at a critical point $\lambda_c$ where, without loss of
generality, we assume there is no symmetry breaking for $\lambda
>\lambda_c$. For the ground state $|\psi (\lambda) \rangle$ in the
symmetric phase, the fidelity $|\langle \psi(\lambda) |g
|\psi(\lambda) \rangle|$ is equal to one for any symmetry operation
$g \in G$. In the broken-symmetry phase $\lambda < \lambda_c$ we can
have
\begin{eqnarray}
0\leq |\langle \psi(\lambda) |g |\psi'(\lambda) \rangle|\leq 1,
\label{broken}
\end{eqnarray}
where $|\psi(\lambda)\rangle$ and $|\psi'(\lambda)\rangle$ are any two states in the degenerate ground state.  This description
is valid for any system admitting a QPT with symmetry-breaking order, regardless of
the type of symmtery group or whether the transitions are continuous or discontinuous.
For later use we will denote the ground-state subspace as $V(\lambda)$.

\begin{figure}
\begin{overpic}[width=81mm,totalheight=50mm]{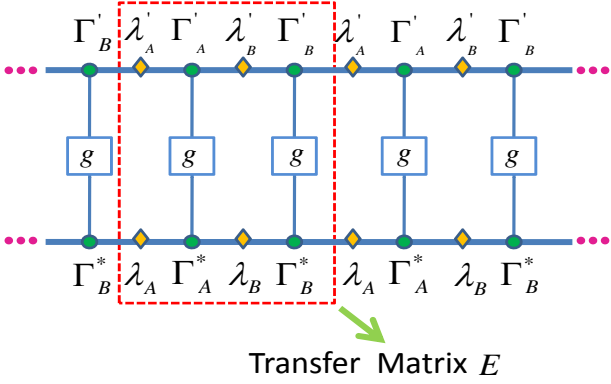}
\end{overpic}
\setlength{\abovecaptionskip}{0pt} \caption{(color online) The
computation of the transfer matrix $E$ from the infinite Matrix
Product State (iMPS) representation of ground state wavefunctions
for translation-invariant quantum system on an infinite
one-dimensional lattice. The transfer matrix $E$ is constructed from
three-index tensors $\Gamma_{A}$, $\Gamma_{B}$, $\Gamma'_{A}$, and
$\Gamma'_{B}$ attached to sites, and  diagonal singular-value
matrices $\lambda_A$, $\lambda_B$, $\lambda'_A$, and $\lambda'_B$
attached to bonds, together with two square gates representing a
nontrivial element $g$ of the symmetry group $G$ acting on physical
indices.} \label{imps}
\end{figure}

Based on this observation we now describe a procedure to construct
order parameters which provide quantitative information about
symmetry-breaking QPTs. Because QPTs generally occur in the infinite
lattice limit, we will utilize an algorithm developed by
Vidal~\cite{vidal}. This algorithm provides an efficient way to
generate an infinite MPS (iMPS) representation of the system's
ground-state wavefunction, and in turn compute expectation values.
Determination of the ground state amounts to computing the imaginary
time evolution operator $\exp(- H \tau)$ acting on an initial state
$|\Psi (0)\rangle $: $ |\Psi (\tau)\rangle = \exp(- H \tau) |\Psi
(0)\rangle / |\exp(- H \tau) |\Psi (0)\rangle |$. In the
symmetry-broken phase with ground-state degeneracy, we can obtain
more than one ground-state representative through different choices
of the initial state $|\Psi (0)\rangle $. Let
$|\psi(\lambda)\rangle$ and $|\psi'(\lambda)\rangle$ denote two such
iMPS representations (which may or may not be equivalent states).
The structure associated to the algorithm, which utlizes the
translational invariance of the system, leads to the conclusion that
\begin{eqnarray}
|\langle \psi(\lambda) |g |\psi'(\lambda)
\rangle|=\lim_{L\rightarrow\infty} |{\rm tr}(E^L)|, \label{overlap}
\end{eqnarray}
where $E$, referred to as the {\it transfer matrix}, is the
four-index tensor schematically defined in Fig.~\ref{imps} in terms
of diagonal singular-value matrices $\lambda_A,\lambda_B$ and the
three-index tensors $\Gamma_A,\,\Gamma_B$. The iMPS representation
becomes exact in the limit as the dimension of the matrices
$\lambda_A,\lambda_B$ approaches infinity. In practice, we introduce
a truncation dimension $\chi$ associated to $\lambda_A,\lambda_B$
and adjust $\chi$ to achieve an extrapolation of results for the
$\chi \rightarrow \infty$ case.

Observe next that (\ref{overlap}) can only take the values zero or one, since the trace is simply the sum of the eigenvalues. This is in stark contrast to (\ref{broken}), and  indicates that the iMPS algorithm is preferential in its convergence to states in the ground-state subspace.  Fixing a representative $|\psi_j(\lambda)\rangle$, we define $W_j(\lambda)={\rm span}\{|\psi_{h,j}(\lambda) \rangle=h|\psi_j(\lambda)\rangle: h\in G\}$. Remembering that the representation of $G$ is unitary, it also follows that the representation of each $g\in G$ on $W_j(\lambda)$ is of the form of a permutation matrix (up to a diagonal unitary transformation). The existence of such a representation for all groups $G$ follows from Cayley's theorem, which states that every group is the subgroup of a symmetric group. It may be that $W_j(\lambda)$ does not span the degenerate subspace of the broken symmetry phase. Within the iMPS algorithm, we can generate subspaces orthogonal to $W_j(\lambda)$ simply by choosing the initial state
$|\Psi (0)\rangle $ of the algorithm to be orthogonal to $W_j(\lambda)$. In principle a basis for $V(\lambda)$ can be constructed in this manner leading to
$\displaystyle V(\lambda)=\oplus_j W_j(\lambda)$. We then still have that the representation of $g\in G$ on $V(\lambda)$, which has a block diagonal structure, is of the form of a permutation matrix.

We can now exploit these facts to uniquely define order parameters which characterize the
broken-symmetry phase. For a fixed choice of $g\in G$ and a fixed iMPS representation $|\psi(\lambda)\rangle$, let $f_g(\lambda)$ denote the square root of the eigenvalue of $E$ which has the largest absolute value.
As such, one sees that $|f_g(\lambda)|=1$ for all
$g \in G$ in the symmetric phase $\lambda >\lambda_c$, and $0\leq |f_g(\lambda)|\leq 1$
in the broken-symmetry phase $\lambda <\lambda_c$.
We now define $I_g(\lambda)$
to be
\begin{equation}\label {fit}
   I_g(\lambda)=\sqrt{1-|f_g(\lambda)|^2}.
 \end{equation}
and argue that the set ${\mathcal O}=\{I_g(\lambda):g\in G\}$ defines a set of order parameters
which detect QPTs with symmetry-breaking order. First, each $I_g(\lambda)$ is  zero if $\lambda >\lambda_c$. Second, each $I_g(\lambda)$ can take a value ranging from 0 to 1 if
$\lambda <\lambda_c$. These features are nothing but what we require
for $I_g(\lambda)$ to be an  order parameter. Moreover, ${\mathcal O}=\{I_g(\lambda):g\in G\}$ is unique on each block representation $W_j(\lambda)\subseteq V(\lambda)$, in that the set is independent of the choice of iMPS representation $|\psi(\lambda)\rangle\in W_j(\lambda)$. To show this, we first consider for each $G$-congugacy class $C$ the set of order parameters ${\mathcal O}_C=\{I_g(\lambda): g\in C\}$. Any different iMPS representation $|\psi_j'(\lambda)\rangle\in W_j(\lambda)$ will be related to $|\psi_j(\lambda)\rangle$ by
$|\psi_j'(\lambda)\rangle=\exp(i\phi)|\psi_{h,j}(\lambda)\rangle$ for some $h\in G$ and some phase $\phi$, due to orthogonality of $|\psi_j(\lambda)\rangle$ and $|\psi_j'(\lambda)\rangle$ as decreed by (\ref{overlap}).
The uniqueness of the set ${\mathcal O}_C$ for $W_j(\lambda)$ follows from
$$ |\langle \psi_j'(\lambda)|g|\psi_j'(\lambda)\rangle| =|\langle \psi_{h,j}(\lambda)|g|\psi_{h,j}(\lambda)\rangle|
= |\langle \psi_j(\lambda)|h^{-1}gh|\psi_{j}(\lambda)\rangle|.$$ Consequently,
${\mathcal O}=\cup_{C} {\mathcal O}_C$ is a uniquely-defined set of order parameters on each subspace block $W_j(\lambda)\subseteq V(\lambda)$. This is in some contrast to the conventional notion of an order parameter, whose value is associated with individual states in $V(\lambda)$.
These considerations are valid for any quantum lattice system where there is
symmetry-breaking order, thus the construction can be applied universally.
Finally, we {\it define} a symmetry-breaking QPT to be second-order if all $I_g(\lambda)$ are continuous functions of $\lambda$. If there are any discontinuous $I_g(\lambda)$ we say the QPT is first-order.

\begin{figure}
\begin{overpic}[width=85mm,totalheight=60mm]{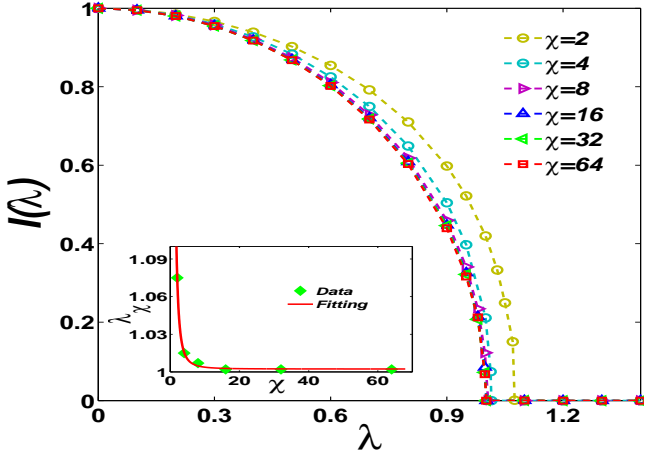}
\end{overpic}
\setlength{\abovecaptionskip}{0pt} \caption{(color online) The
universal order parameter $I(\lambda)$ for the quantum Ising model
in a transverse magnetic field.
For the control parameter $\lambda$ less than the critical value
$\lambda_c=1$, the $\mathbb{Z}_2$ symmetry is spontaneously broken. In the
iMPS simulation, a transition point $\lambda_{\chi}$
occurs for a given truncation dimension $\chi$. As the truncation
dimension $\chi$ is increased, $\lambda_{\chi}$ approaches
the exact value. Note that the factorized field $\lambda_f$ occurs
at $\lambda_f =0$, at which the universal order parameter takes the
maximum value. Inset: the critical point $\lambda_{c}$ is determined
from an extrapolation of the pseudo phase transition point
$\lambda_{\chi}$ with respect to the truncation dimension $\chi$.
Here, the fitting function is $\lambda_{\chi} = \lambda_{c} +
a\chi^{-b}$, with $\lambda_{c} = 1.0023$,\;$ a = 0.3937$ and $b =
2.4390$. } \label{ising}
\end{figure}

\begin{figure}
\begin{overpic}[width=85mm,totalheight=60mm]{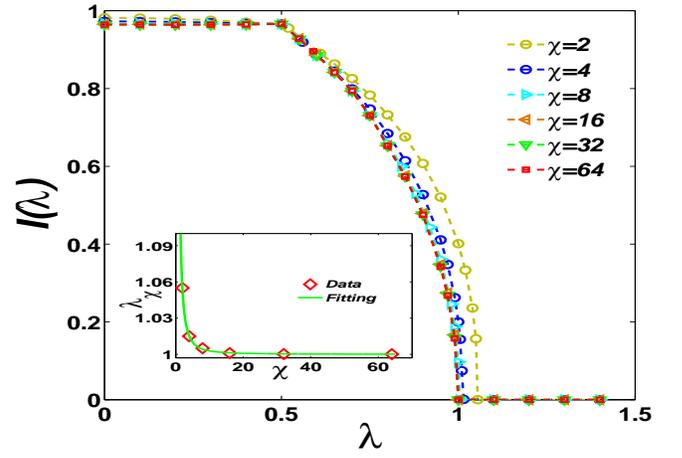}
\end{overpic}
\setlength{\abovecaptionskip}{0pt} \caption{(color online) The
universal order parameter $I(\lambda)$ for the quantum XY model in a
transverse magnetic field.  For increasing $\lambda$, a
transition point $\lambda_{\chi}$ occurs as
$I(\lambda)$ goes to zero. With increasing truncation
dimension $\chi$, the transition point
$\lambda_{\chi}$ approaches the critical point $\lambda_{c}=1$. The universal order
parameter $I(\lambda)$ reaches the maximum value when $\lambda_{f}$
is equal to 0.5. Inset:
the critical point $\lambda_{c}$ is determined from an extrapolation
of the pseudo phase transition point $\lambda_{\chi}$ with respect
to the truncation dimension $\chi$. Here, the fitting function is
$\lambda_{\chi} = \lambda_{c} + a\chi^{-b}$, with $\lambda_{c} =
1.000039$,\;$ a = 0.197301$ and $b = 1.845012$. }\label{xy}
\end{figure}

\begin{figure}
\begin{overpic}[width=85mm,totalheight=60mm]{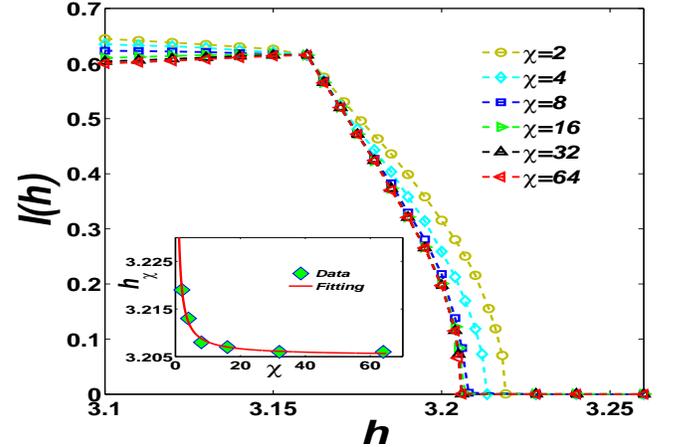}
\end{overpic}
\setlength{\abovecaptionskip}{0pt} \caption{(color online) The
universal order parameter $I(h)$ for the spin-1/2 quantum XYX model
in an external magnetic field. A
transition point $h_{\chi}$ occurs as as $I(h)$
vanishes. As the truncation dimension $\chi$ increases, the transition point $h_{\chi}$ approaches the critical point
$h_c$. In addition, the universal order parameter $I(h)$ takes the
maximum value at $h_{f} \sim 3.16$. Inset: the critical point
$h_{c}$ is determined from an extrapolation of the pseudo phase
transition points $h_{\chi}$ with respect to the truncation
dimension $\chi$. Here, the fitting function is $h_{\chi} = h_{c} +
a\chi^{-b}$, with $h_{c} = 3.2052$,\;$ a = 0.0277$ and $b = 0.9848$.
} \label{xyx}
\end{figure}

\begin{figure}
\begin{overpic}[width=85mm,totalheight=60mm]{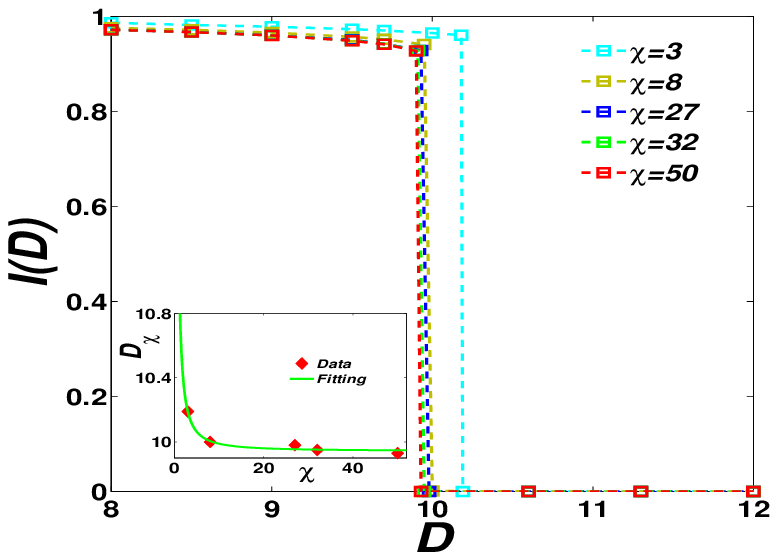}
\end{overpic}
\setlength{\abovecaptionskip}{0pt} \caption{(color online) The
universal order parameter $I(D)$ for the quantum spin-1 XXZ model
with single-ion anisotropy, which exhibits discontinuity at the
transition point $D_{\chi}$. This indicates that a first-order QPT
occurs at $D=D_{\chi}$. As the truncation dimension $\chi$
increases, the transition point $D_{\chi}$ approaches the critical
point $D_c = 9.9434$. Inset: the fitting function is $D_{\chi} =
D_{c} + a\chi^{-b}$, with $D_{c} = 9.9434$,\;$ a = 1.1490$ and $b =
1.4033$. } \label{xxz}
\end{figure}

{\it The models.}   To illustrate our scheme, we consider
three one-dimensional lattice Hamiltonians with ${\mathbb Z}_2$-symmetry.
This group is generated by a single non-trivial element $g$, which squares to the identity, and in each case admits a unitary representation.
The construction described above thus leads to a single order parameter, which we hereafter denote as $I(\lambda)$.

The first of the models is the quantum XY model in a transverse
magnetic field in an infinite-size lattice in one spatial dimension.
The Hamiltonian takes the form,
\begin{equation}\label {fitxy}
   H = - \sum_{i=-\infty}^{\infty}
   \left(\frac{1+\gamma}{2} S_{x}^{[i]}S_{x}^{[i+1]}+\frac{1-\gamma}{2} S_{y}^{[i]}S_{y}^{[i+1]}+\lambda
   S_{z}^{[i]}\right),
 \end{equation}
where $S_{\alpha}^{[i]} \; (\alpha = x, y, z)$ are the spin-1/2
Pauli operators at lattice site $i$,  $\lambda$ is a transverse
magnetic field, and $\gamma$ is an anisotropic coupling constant.
The model is invariant under the symmetry operation: $S_{x}^{[i]}
\rightarrow -S_{x}^{[i]}$, $S_{y}^{[i]} \rightarrow -S_{y}^{[i]}$
and $S_{z}^{[i]} \rightarrow S_{z}^{[i]}$ for all sites, which
yields the $\mathbb{Z}_{2}$ symmetry. For nonzero $\gamma$, it is critical at
$\lambda =1$~\cite{cris}. In addition, a factorizing field occurs at
$\lambda_f = \sqrt {1-\gamma^2}$. If $\gamma=1$, the quantum XY
model reduces to the quantum Ising model in a transverse field.

We also investigate the quantum spin-1/2 XYX model in an external
magnetic field. The Hamiltonian can be written as
\begin{equation}\label {fitxyx}
   H = \sum_{i=-\infty}^{\infty} \left(S_{x}^{[i]} S_{x}^{[i+1]} + \Delta_{y}S_{y}^{[i]} S_{y}^{[i+1]} + S_{z}^{[i]} S_{z}^{[i+1]} + hS_{z}^{[i]}\right),
 \end{equation}
where $S_{\alpha}^{[i]} \; (\alpha = x,y,z)$ are the Pauli spin
operators at site $i$, $\Delta_{y}$ is a parameter describing the
rotational anisotropy, and $h$ is an external magnetic
field. This model also possesses a $\mathbb{Z}_{2}$ symmetry, with the
symmetry operation: $S_{x}^{[i]} \rightarrow -S_{x}^{[i]}$,
$S_{y}^{[i]} \rightarrow -S_{y}^{[i]}$ and $S_{z}^{[i]} \rightarrow
S_{z}^{[i]}$ for all sites. Below we shall choose $\Delta_{y} =
0.25$. In this case, the critical magnetic field is $h_{c} \sim
3.210(6)$~\cite{mc}, with  a factorizing field $h_{f} \sim 3.162$.

The last model considered here is the quantum spin 1 XXZ model with
single-ion anisotropy.  The Hamiltonian takes the form,
\begin{equation}\label {fitxxzd}
   H = \sum_{i=-\infty}^{\infty} \left[J(S_{x}^{[i]} S_{x}^{[i+1]} + S_{y}^{[i]} S_{y}^{[i+1]}) +J_{z}S_{z}^{[i]} S_{z}^{[i+1]}\right] +
    D\sum_{i=-\infty}^{\infty}S_{z}^{[i]2},
 \end{equation}
where $S_{\alpha}^{[i]} \; (\alpha = x,y,z)$ are the spin-1
operators at the $i$-th lattice site , $D$ represents uniaxial
single-ion anisotropy. We choose $J = 1$, $J_{z} = 10$, with $D$ as
the control parameter. As such, the system undergoes a first-order
QPT from a gapped $\mathbb{Z}_2$ symmetry-broken  N\'eel phase to a gapped
$\mathbb{Z}_{2}$ symmetric large-$D$ phase, with the symmetry operation:
$S_{x}^{[i]} \rightarrow S_{x}^{[i]}$, $S_{y}^{[i]} \rightarrow
-S_{y}^{[i]}$ and $S_{z}^{[i]} \rightarrow -S_{z}^{[i]}$ for all
sites.

{\it The results.}
In Fig.\ref{ising}, we plot the universal order parameter
$I(\lambda)$ for the quantum Ising model in a transverse field, with
the field strength $\lambda$ as the control parameter. For
the control parameter $\lambda$ less than a pseudo critical value $
\lambda_{\chi}$, the universal order parameter $I(\lambda)$ is
non-zero, which characterizes the $\mathbb{Z}_{2}$ symmetry-broken phase. In
the symmetric phase $\lambda
> \lambda_{\chi}$, the universal order parameter $I(\lambda)$ is zero.
When the control parameter $\lambda$ varies across the point $\lambda_\chi$, the behavior of the universal order
parameter $I(\lambda)$ implies that the system undergoes a phase
transition at the $\lambda_{\chi}$. As the
truncation dimension $\chi$ is increased, $\lambda_{\chi}$ moves toward the known critical point
$\lambda_c=1$. Performing an extrapolation of
$\lambda_{\chi}$ with respect to $\chi$, we  obtain
$\lambda_{c}=1.0023$. Note that the universal order parameter
reaches the maximum value at $\lambda = 0$. The maximum value of $I(\lambda)$ coincides with the existence of a
factorized state at this point, in which no entanglement
exists.

In Fig.\ref{xy}, the universal order parameter $I(\lambda)$ for the
quantum spin 1/2 XY model in an external magnetic field $\lambda$ is
plotted, with the magnetic field strength $\lambda$ as the control
parameter. A transition point $\lambda_{\chi}$ occurs as
$I(\lambda)$ varies from zero to nonzero values.  With increasing
trucation dimension the transition point $\lambda_{\chi}$ approaches
the critical point $\lambda_{c} = 1$. Performing an extrapolation
with respect to $\chi$ yields the critical point $\lambda_{c} =
1.000039$. A factorizing field $\lambda_f$ occurs at $\lambda_{f}
=0.5$, where again $I(\lambda)$ takes its maximum value.

In Fig.\ref{xyx}, we show the universal order parameter $I(h)$
between a ground state and its symmetry-transformed counterpart for
the quantum XYX model in an external magnetic field. In the range $h
< h_{\chi}$, the universal order parameter $I(h)$ is non-zero, which
characterizes the $\mathbb{Z}_{2}$ symmetry-broken phase. A
transition point $h_{\chi}$ occurs as $I(h)$ vanishes.  As the
truncation dimension $\chi$ increases, the transition point
$h_{\chi}$ approaches the critical point $h_c = 3.2052$.  Performing
an extrapolation of $h_{\chi}$ with respect to $\chi$, we  obtain
$h_{c}=3.2052$, $a=0.0277$, $b=0.9848$. The universal order
parameter $I(h)$ takes the maximum value at $h_{f} \sim 3.16$.

In Fig.\ref{xxz}, the universal order parameter $I(D)$ is plotted
for the quantum XXZ model with single-ion anisotropy. For $D <
D_{\chi}$, the system is in the gapped $\mathbb{Z}_2$
symmetry-broken N\'eel phase, so the universal order parameter
$I(D)$ is nonzero.  For $D > D_{\chi}$, it is in the gapped
$\mathbb{Z}_{2}$ symmetric large-$D$ phase, so $I(D)$ is zero. We
see that $I(D)$ is discontinuous, providing an example of a
first-order QPT. As the truncation dimension $\chi$ increases, the
transition point $D_{\chi}$ approaches the critical point $D_c =
9.9434$.

{\it Summary.} We have described a universal procedure to compute
order parameters for any translation-invariant quantum lattice
system with a symmetry group $G$. The scheme has been illustrated
for the quantum Ising model in a transverse magnetic field, the
quantum spin-1/2 XYX model in an external magnetic field, and the
quantum spin-1 XXZ model with single-ion anisotropy,  by exploiting
the infinite MPS algorithm for one-dimensional systems. In all
instances the procedure is successful in identifying the QPT points.
Moreover, we observe that occurences of factorizing fields are in
correspondence with the maximal values of the order parameters.

{\it Acknowledgements.} We thank John Fjaerestad and Guifre Vidal
for insightful comments. This work is supported in part by the
National Natural Science Foundation of China (Grant Nos: 10774197
and 10874252), the Natural Science Foundation of Chongqing (Grant
No: CSTC, 2008BC2023).


\end{document}